\documentclass[aps,twocolumn,showpacs]{revtex4}
\usepackage{graphicx}
\usepackage{amsmath}
\usepackage{times}
\begin{document}
\def\T{\Theta}
\def\D{\Delta}
\def\d{\delta}
\def\r{\rho}
\def\p{\pi}
\def\a{\alpha}
\def\g{\gamma}
\def\ra{\rightarrow}
\def\s{\sigma}
\def\b{\beta}
\def\e{\epsilon}
\def\G{\Gamma}
\def\om {\omega}
\def\pe{$1/r^\a$ }
\def\l{\lambda}
\def\f{\phi}
\def\w{\psi}
\def\m{\mu}
\def\t{\tau}
\title{Binaries and core-ring structures in self-gravitating systems}
\author{I.~Ispolatov}
\affiliation{Departamento de Fisica, Universidad de Santiago de Chile,
Casilla 302, Correo 2, Santiago, Chile}
\date{\today}
\begin{abstract}
Low energy states of self-gravitating
systems with finite angular momentum are considered.
A constraint is introduced to confine cores and other condensed objects
within the system boundaries by gravity alone. 
This excludes previously observed astrophysically irrelevant
asymmetric configurations with a single core.
We show that
for an intermediate range of a short-distance cutoff and small
angular momentum, the equilibrium configuration is
an asymmetric binary. 
For larger angular momentum  or for a smaller range of the 
short distance cutoff, 
the equilibrium configuration consists of a central core
and an equatorial ring.
The mass of the ring
varies between zero for vanishing rotation
and the full system mass for the maximum angular momentum $L_{max}$
a localized gravitationally bound system can have.
The value of $L_{max}$ scales as $\sqrt{\ln(1/x_0)}$,
where $x_0$ is a ratio of a short-distance 
cutoff range to the
system size. An example of the soft gravitational potential is 
considered; the conclusions  
are shown to be valid
for other forms of short-distance regularization.
\end{abstract}

\maketitle

\section{introduction}
Despite an abundance of rotating structures in the Universe,
the statistical mechanics of these systems is not completely
understood. The non-rotating self-gravitating systems,
as shown by  the Mean Field (MF) analysis and
recently confirmed by direct computer simulations, 
exhibit two phases (see, for example, \cite{pr,dv,chi,us1} and
references therein): A
high energy ``uniform'' phase where density contrast is small,
and a low energy phase consisting of a diluted halo 
and a dense core. The structure of the core is determined by
a type of a short-range regularization, which can vary from
the exclusion in phase or coordinate space to a small-distance truncation
of the interaction potential.
At the same time,
the structure 
of rotating self-gravitating systems, particularly the form and number of
cores or other dense objects
in the low energy states, remains a subject of discussion.
Configurations such as symmetric and asymmetric binaries and rings 
\cite{vot,vot1,vot2,vot3}, a single spheroidal core with or without
an equatorial ring \cite{ch1, ch2}, and a single ``asymmetric'' core, sliding
along the container wall \cite{us},
have been found in rotating self-gravitating systems. 
Among possible reasons for this plethora of equilibrium
configurations the boundary conditions play definitely an important role.
As in the non-rotating case (see, for example, \cite{pr}), a confining 
box with reflecting
walls (to conserve the  
energy and mass) is essential for the existence of any equilibrium state. 
For a rotating system, it is natural to require angular momentum
conservation as well and
chose an axisymmetric boundary; usually a spherical container is considered 
\cite{vot,ch2,us,la}. However, the reflective boundary conditions seem to be  
the main reason for 
the apparent discrepancy 
between the particle simulations and the MF results:
While a true particle system inevitably evolves towards an asymmetric
single-core  configuration
\cite {us}, 
the states with an arbitrary number of cores and even rings are predicted by
the MF analysis
\cite{vot,vot1,vot2,vot3,ch2}. To obtain two- and multi-core states, 
the center of mass of the system
is fixed in the center of the
container \cite{vot,vot1,vot2, vot3,votr}. While in the MF analysis
the center of mass constraint is implemented by
removing the dipole terms from the multipole expansions of density and
potential, it is not clear  
how to  fix the center of mass
in a particle simulation. The total momentum 
and the
center of mass position, unlike
the angular momentum, are not conserved even in a spherical system, 
when the boundaries are reflecting. 
Due to the ergodicity of a 3-dimensional Coulomb system,
even a specially prepared, highly symmetric initial state (with a
a centrally-symmetric coordinates and momenta distribution) 
will evolve
towards the usually non-symmetric highest entropy configuration.
For a rotating system, 
the most probable configuration consists of the single core
sliding along the container wall
\cite{us,votr}. 
The reason why the state with a single core has the highest entropy (or the
lowest 
free energy)  is the following: When two or more initially separated cores
merge, the gravitational potential energy decreases, which, for the fixed
total energy, leads to a gain in 
the translational entropy of the halo.
Intuitively, as there 
is no naturally occurring analog of a container wall on which the core
may slide, this asymmetric state with a single core looks highly unphysical. 

If particle simulations seem unable to reproduce
the configurations obtained by the MF methods and exhibit only
a physically-irrelevant state with a single core,
is there any other way to 
validate the MF results? 
The criteria of physical relevance, which the
asymmetric state with a single core does not satisfy, are rather intuitive and 
can be formulated as follows:
The physically relevant equilibrium states of rotating systems must be
affected by the boundary conditions in the least possible way.
This minimal boundary condition effect is attained in the case of 
the core-halo states of non-rotating self-gravitating
systems: If a container surrounding such system is removed, the halo will
start to evaporate, while the core will remain almost intact 
for a considerable time (see \cite{us2} for an estimate of core-halo
thermalization rate). On the 
contrary, the rotating asymmetric state with a single
core  
would undergo significantly more dramatic changes if the
container were removed: The core, being no longer supported by the container
wall,  will
escape ballistically with nothing left within the system
boundaries. 
Similarly to the non-rotating case, the
minimal boundary condition effect can be attained
in rotating systems if cores (or other condensed objects such as disks or
rings) are confined within the system
boundary by gravity alone. 
In the limit of the ground-state energy 
(or zero temperature), 
such a state will consist only of the gravitationally bound
condensed parts, unaffected  
by the removal of the container at all. 

In this paper we consider such weakly interacting with container,
or ``physically relevant'' states
of rotating 
self-gravitating systems. To satisfy the relevance criteria suggested
above, one needs to
find core orbits in the presence of a halo. 
To simplify this task, we consider the limit of ground-state  energy 
(or zero temperature) in which the gaseous halo is 
condensed into the cores, and
there is no internal motion of core particles. 
The only motion in this limit is the macroscopic movement
of cores, specified by the angular momentum constraint.
For sufficiently low energies, 
the state 
with the lowest
ground-state energy is the thermodynamically equilibrium one
(while the other mechanically stable states are thermodynamically 
metastable).
Thus, to determine a core structure of the physically relevant
low-energy equilibrium state, it is sufficient to find
a gravitationally bound core configuration with the lowest energy of a given
mass, size, and
angular momentum.

The number of candidates
for the lowest energy core state can be significantly reduced 
using the following heuristic argument.
The energy of an ensemble of rotating cores is minimized when the mass is 
concentrated into 
the largest core.
This is so because for sufficiently small short-range cutoff,
the total energy of a rotating self-gravitating 
system is dominated by the 
negative gravitational self-energy of the cores. The absolute value of 
gravitational 
self-energy of a core grows with the core mass faster than linearly for all
reasonable forms of short-range cutoff. Hence,
the 
energy is minimized by a configuration which consists of the principal core 
of the maximum possible mass,
while the remaining mass is distributed to carry 
the given angular momentum in the most
energy-efficient way.
For this reason the particle system in simulations always evolves towards the
asymmetric configuration consisting of the single core which carries all
the angular momentum itself \cite{us}. Without a constraint that the
system has to 
occupy 
a finite volume, the lowest-energy state would have consisted of a core
containing all but one particle with that particle having an orbit 
radius defined by the angular momentum constraint. 
Likewise, with the spatial localization constraint,
the most mass- and energy-efficient way to carry the given angular
momentum is to put the smallest possible mass on a circular orbit of the
maximum allowed radius. Being evident for a two-body system 
(see, for example, \cite{ll}), the energy efficiency of circular orbits 
can be seen from the following argument: If a circular orbit
is perturbed by adding a radial component to the velocity, 
the angular momentum is unchanged while the energy increases.
Below we consider two possible 
configurations consisting of 
the principal core and
the remaining mass on a single circular orbit:
A central core with a ring of $N$ orbiting cores
and a binary, generally asymmetric. A symmetric binary is a limiting case
of both families. Examples are abundant in
the Universe and have been
observed in the MF analysis \cite{vot, vot2}
as possible equilibrium or metastable configurations. 
Other mechanically stable
rotating core configurations that do not belong to these two families
apparently have less mass concentrated into the largest core, and thus have
higher energy.

Using simple mechanics, we will derive that the choice of the lowest
energy state depends on the range of a short-range
(or high-density) regularization and the angular momentum. 
For an intermediate range of the
small-distance regularization and small angular momentum, 
the binary state has the lowest ground-state
energy, which
confirms the results of \cite{vot3}. In a limit of the vanishing range
of the cutoff, or for the higher angular momentum,
the core-ring state becomes the equilibrium one.
The paper is organized as follows:
after this introduction we define the model more formally. In Section III
we compare the ground-state energies of two families
of systems: A central core with a multicore ring, and an asymmetric binary.
A Discussion and Conclusion section completes the paper.

\section{Definition of the model}
Let 
us now formally define the model.
We search for the lowest-energy state of a 
system of $M \gg 1$ self-gravitating unit mass particles
with fixed total angular momentum $L$ localized within a sphere of radius $R$.
To make the problem analytically tractable, we will limit our consideration
to the case when the range of small-distance (high-density) 
regularization is short. Hence the 
volume of all condensed objects in the ground state 
is considered to be negligible compared to the volume of the
system.
In addition, to make a clear distinction from the non-rotating case, we focus
our attention on the sufficiently high values of 
angular momentum $L$ to exclude 
the configuration with the single spinning central core. 
Most of the analysis below is for a system
of classical particles interacting via the attractive soft Coulomb 
potential $-(r^2+r_0^2)^{-1/2}$ (the gravitational constant is set to be
unity). This simple form of short-range regularization
is qualitatively equivalent to
other forms of ``softening'', such as truncation of the Fourier
or spherical harmonic expansions. For an interparticle distance $r$ smaller
than the respective softening radius
(given by $r_0$, or by a wavelength of the highest
untruncated harmonic function)
all  soft potentials tend to a harmonic oscillator
potential. Consequently, for all soft potentials
a condensed core with no particle motion (at 
zero temperature) is a point-like object. 
This is different from, for example,  
the ground state of a system of fermions or 
hard core particles with the finite core volume. 
Qualitative arguments will be given to show that conclusions made for the
system with soft potential also hold
for systems with other forms of short-range regularization.

In addition to the notations 
$r$, $E$, and $L$ for distance, energy, and absolute
value of angular momentum, in the following we will also use the
rescaled (universal) units for 
distance $x$, energy $\e$, and 
angular momentum $\l$ (see, for example, \cite{pr,chi,vot}).
\begin{eqnarray}
\label{def}
\nonumber
x & \equiv & \frac{r}{R}\\
\e & \equiv &\frac{ER}{M^2}\\
\nonumber
\l  & \equiv & \frac{L}{\sqrt{M^3R}}.
\end{eqnarray}

\section{core-ring and asymmetric binaries}

First consider a ring, i.e.
a system of $N$ identical cores (point masses in the case of a 
soft potential)
of mass $M/N$ each 
moving on a circular orbit of radius $r\leq R$. 
Assume that the 
softening radius $r_0$ 
is much smaller than the distance between the neighboring
cores, and the interaction between cores can be
considered as 
bare gravitational one.

Mechanical equilibrium requires the centripetal acceleration of each
core to be equal to the total force exerted on it by the others,
\begin{equation}
\label{om}
\om^2 r= \sum_{i=1}^{N-1}\frac{M \cos[(\pi-\phi_i)/2]}{2 N r^2(1-\cos{\phi_i})}
\end{equation}
with $\phi_i=2\pi i/N$. This gives the 
angular momentum of the system 
\begin{equation}
\label{L}
L^2=M^3r\frac{1}{4N}\sum_{i=1}^{N-1}\frac{1}{\sin(\pi i/N)}
\mathop
{\longrightarrow}_{N \to \infty}
M^3r\frac{\ln N + C}{2\pi},
\end{equation} 
where for large $N$ the sum is replaced by an integral.
This asymptotic expression turns out to be quite robust
(less than 4\% off, even for $N=2$), and the small numerical value of
the constant $C \approx 0.126$ makes $C$ negligible even for few-core systems.

It follows from (\ref{om},\ref{L}) 
that a binary system ($N=2$) localized within a sphere of
radius $R$ cannot have rescaled angular momentum higher than 
\begin{equation}
\label{Lbin}
\l_{bin}=1/\sqrt{8}\approx 0.35.
\end{equation}
In a ring, 
the number of ring cores 
grows as  $N \sim \exp(2\pi\l^2)$ for large $\l$.
The increase of $N$, however, cannot continue indefinitely:
For the interaction between cores to be similar to the bare gravitational 
potential $-1/r$,
the distance between cores should be larger than the
softening radius $r_0$. Hence, when the number of cores exceeds the
corresponding limit $N>N_{max}\approx 2 \pi r/r_0$, 
the cores may still get closer to each other,
while the maximum force acting on each core saturates at the corresponding
to $N_{max}$ value.
Consequently, the maximum angular momentum
of a self-gravitating system with 
radius not greater than $R$ is  
\begin{equation}
\label{Lmax}
\l_{max}\approx \sqrt{\frac{\ln (R/r_0)}{2\pi}}=
\sqrt{\frac{|\ln x_0|}{2\pi}}.
\end{equation} 
Qualitatively similar estimate for the maximum angular momentum 
$\l_{max}$ 
exists for systems with finite-size cores, such as those formed by 
ensembles of fermions or hard-core particles. For such systems, 
the maximum number of cores $N_{max}$ corresponds to
a merging of cores into a continuous (finite-volume) ring or torus. 
The radius of the body of the torus $r_c$
(which is of the order of the radii of cores before merging),
serves the role of a cutoff parameter $r_0$ in (\ref{Lmax}).
This conclusion can also be reached 
using the following argument: One can split a continuous 
ring into $N_{max}$ segments of size
$\sim r_c$ and consider the interaction between them in the multipole
expansion. 
The monopole-monopole interaction 
gives rise to the leading logarithmic term
in (\ref{L}, \ref{Lmax}), while the higher-order terms produce only
${\cal O}(r_c/R)^0$ corrections.

\begin{figure}
\includegraphics[width=.2\textwidth, height=.2\textwidth]{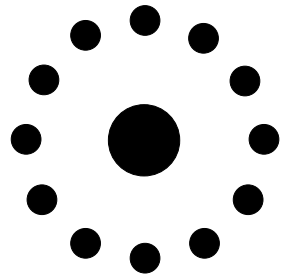}
\caption{\label{fig1} A sketch of a central core -- ring system.
} 
\end{figure}
Let us now consider a core-ring system consisting of a single central core of 
mass $M(1-\a)$
and a ring of mass $M\a$ consisting of $N\geq 2$ cores. 
This structure, sketched in Fig.~\ref{fig1},
resembles the planet Saturn with its 
ring. 

Similarly to the Eq.~(\ref{L}), the equation of motion for an orbiting core
yields for the angular momentum 
of the system:
\begin{equation}
\label{Ls}
L^2= M^3 r \a^2\left [\a \frac{\ln N}{2\pi}+ (1-\a)\right ].
\end{equation} 
The total energy of the core-ring system consists of the
gravitational self-energies of the central core and orbiting cores, and the 
energy of macroscopic rotation:
\begin{equation}
\label{sate}
\e =  -  \frac{(1-\a)^2}{2x_0}  - \frac{\a^2}{2Nx_0}
- \frac{\a^3[\a({\ln N}/(2\pi)-1]+1]^2}{2\l^2}.
\end{equation} 
To find the ground state energy, this expressions must be minimized with
respect to $\a$ and $N$, $2\leq N\leq 1/x_0$, taking into account 
the $r\leq R$ constraint which has the form
\begin{equation}
\label{satc}
\a^2\left (1-\a+ \a \frac{\ln N}{2\pi} \right) \geq \l^2.
\end{equation} 
For sufficiently small $x_0$, the first term in (\ref{sate}) is the dominant
one. Hence, the minimum energy is reached  when $N$ is increased to its 
saturation value 
$N \sim 1/x_0$ to minimize the relative 
ring mass $\a$ within the range allowed by
(\ref{satc}). 
\begin{figure}
\includegraphics[width=.45\textwidth, height=.45\textwidth]{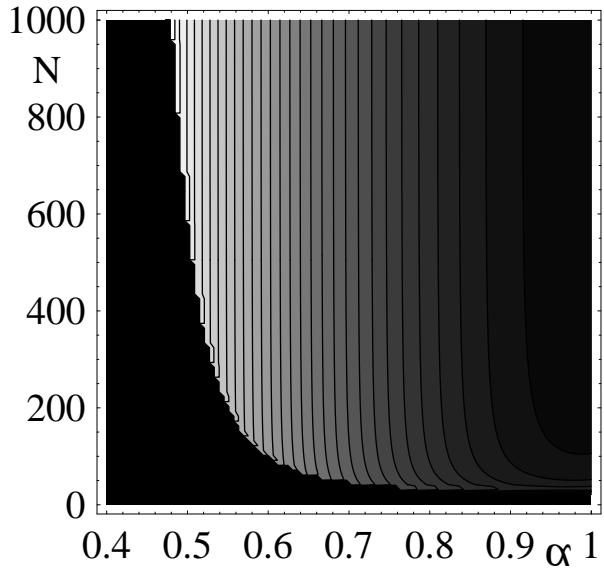}
\caption{\label{fig2} Contour plot of $\e$ vs $\a$ and $N$, 
defined by the Eq.~(\ref{sate}) and 
constraint (\ref{satc}) for $\l=0.5$ and $x_0=10^{-3}$.
The energy decreases (increases by absolute value) 
from dark to light, and reaches its minimum at the largest possible
$N$, $N=1/x_0=1000$ and
the smallest $a$ allowed by (\ref{satc}), $\a \approx 0.49$. 
The black area
corresponds to values of $\a$ and $N$ which do not satisfy the localization
constraint $r\leq R$ (\ref{satc}). 
} 
\end{figure}
This is illustrated by the example presented in Fig.~\ref{fig2}.
The ground state energy of the core-ring configuration 
with the angular momentum
$\l\leq \ln N/2\pi$ is 
\begin{equation}
\label{e_cr}
\e_{c-r} =  -  \frac{(1-\a)^2}{2x_0}  - \frac{\a^2}{2}
- \frac{\a}{2} \left( \frac {\a |\ln x_0|}{2\pi}+1 - \a \right),
\end{equation} 
where  $\a(\l)$ is the minimal ring mass allowed by (\ref{satc}) and $N=1/x_0$.

The mass distribution of the 
ground states of the core-ring systems with other forms of cutoff 
is similar.
The self-energy of a spherical core of $m$ particles can be expressed as
$E_{self}^{HC}= - C_{HC}m^{5/3}$ and 
$E_{self}^{F}= -C_{F}m^{7/3}$ for systems of hard-core particles
and fermions, respectively (see \cite {chandr} for the energy of
self-gravitating fermion ball). The constants $C_{HC}$ and $C_F$, which 
depend on the hard-core radius and the number of internal degrees of freedom, 
play the
role of $1/r_0$ in (\ref{e_cr}). For
sufficiently large values of these constants (or equivalently, sufficiently
small cores), the self-energy of
the central core dominates over the self-energies of the ring cores and the 
energy of macroscopic motion. Hence, for a reasonable form of the 
short-range cutoff, 
the lowest energy core-ring configuration consists
of the central core of the largest possible mass allowed by the localization
constraint, and the orbiting ring
of the system radius $R$ which carries all the angular momentum. 

Now let us consider a binary, generally asymmetric \cite{vot3}, as
another candidate for the lowest energy configuration with a
moderate angular momentum $\l<\l_{bin}$ (\ref{Lbin}).
Introducing an asymmetry parameter 
$0 < \delta< 1/2 $ and the core separation $r\leq 2R$, 
denote the core masses and orbit radii  
as
$M_1=\delta M$, $M_2=(1- \delta) M$, $r_1= (1- \delta) r$, and 
$r_2=  \delta r$, correspondingly. 
The total energy consists of the energy of
rotation and the gravitational self-energy of the cores, and
reads in rescaled units:
\begin{equation}
\label{as}
\e=  -  \frac{[\d(1-\d)]^3}{2\l^2} - \frac{(1-\d)^2+\d^2}{2x_0}.
\end{equation} 
To find the ground-state energy, this expression has to be minimized with
respect to $\d$ while taking into account the
constraint that the radius of the largest orbit $r_1$ should not exceed $R$,
\begin{equation}
\label{asc}
\d^2(1-\d)\geq \l^2.
\end{equation} 
For all reasonable values of $x_0$ ($x_0<32\l^2/3)$ the
gravitational self-energies of the cores dominates the 
total energy $\e$ in Eq.~(\ref{as}). 
Hence, the
energy is minimized
\begin{equation}
\label{asmin}
\e_{bin}=  -  \frac{\d(1-\d)^2}{2} - \frac{(1-\d)^2+\d^2}{2x_0}
\end{equation} 
with the highest possible asymmetry, or
the smallest possible value of $\d \leq 1/2$ allowed by (\ref{asc}).
The same is true for other forms of short-range regularization:
Since the self-energy of a self-gravitating core grows faster than
linearly with the core mass, a binary with the highest possible asymmetry will
always have the lowest energy.
 
This confirms the conclusion made in \cite{vot3} that
for low energy an asymmetric binary has a higher entropy than
a symmetric one, albeit with the restriction that orbits in a binary are  
gravitationally-supported only for $\l < \l_{bin}$.
 
\begin{figure}
\includegraphics[width=.45\textwidth, height=.45\textwidth]{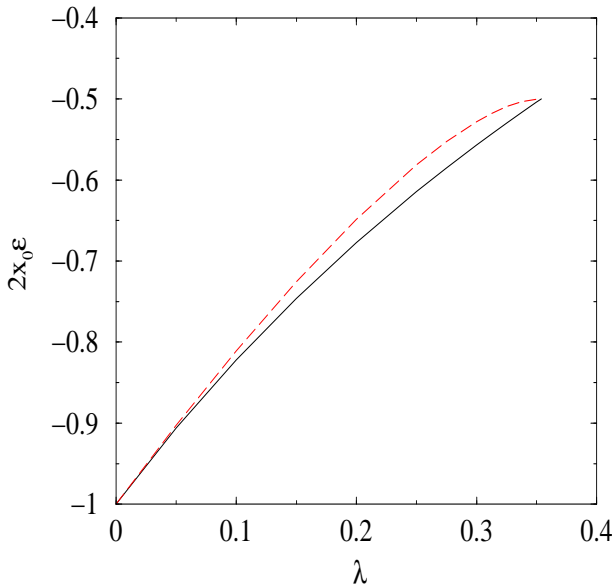}
\caption{\label{fig3} Plot of the rescaled ground state energies 
of core-ring
$\e_{c-r}$ (solid line) and  asymmetric binary $e_{b}$ (dashed line)
configurations  
vs angular momentum $\l$ for $x_0=10^{-7}$. 
} 
\end{figure}

To compare the ground state energies of core-halo and asymmetric binary
configurations,  the cubic equations (\ref{satc},\ref{asc}) have to be solved
for a given angular momentum $\l\leq \l_{bin}$, 
and the resulting
$\a$ and $\d$ are to be substituted 
into the expressions for  energy (\ref{e_cr},\ref{asmin}).
For sufficiently small cutoff radius $x_0 \ll 1$, the self-energies of the
cores, 
described by the first terms in
(\ref{e_cr},\ref{asmin}), give dominant contributions to the total energy.
Qualitatively,
when the cutoff radius is still not too small, so that $|\ln x_0|/2 \pi \ll
1$, the logarithmic term in (\ref{satc}) is negligible and $\a \approx \d$.
In this case, due to
the noticeable contribution of the self-energy of
the smaller core,
$-\d^2/2x_0$,
the binary system has the lowest energy.  
On the other hand, if $|\ln x_0|/2 \pi \gg 1$, the 
logarithmic term dominates in (\ref{satc}) and   $\a < \d$.
In this case, the self-energy of the bigger central core of the core-ring
system becomes smaller than the self-energies of both binary cores, and the
core-ring system is the equilibrium one.
The $\l$-dependence of the binary and core-halo energies in the 
borderline case of $x_0 \approx 10^{-7}$ is illustrated in Fig.~\ref{fig3}.
Hence for $x_0<10^{-7}$, the core-ring configuration is the equilibrium one
for all 
values of angular momentum $\l$, while for  $x_0>10^{-7}$ there exists a range
of $\l\leq \l_{bin}$ for which the binary system is the equilibrium one.
Naturally, the energies of the core-ring and binary configuration 
coincide for $\l=0$ where both configurations are reduced to just one central
core. 

\section{Discussion and Conclusion}
In the previous sections we considered the low-energy equilibrium states of
rotating self-gravitating systems and arrived at the following conclusions:

To set a distinction between the physically relevant and irrelevant states, we
suggested  that the core part of the relevant states has to be moving on
gravitationally supported orbits and not come in contact with the system
boundary. 
In the low-energy or zero-temperature limit, such equilibrium states 
become localized core-only states and can exist without a container.

We found that, depending on the range of the short-distance regularization and
the angular momentum, two
possibilities exist for the equilibrium
configurations of rotating self-gravitating systems: 
For an intermediate range of short-distance cutoff and small
angular momentum $L
\leq \sqrt{M^3R/8}$, 
an asymmetric two-core binary configuration has the lowest ground-state energy
and thus is the equilibrium one \cite{vot3}. 
For  $L > \sqrt{M^3R/8}$, or for a very small range of the cutoff
and 
arbitrary angular
momentum, the equilibrium configuration consists of a central core
and an equatorial ring.

The precise value of the cutoff range at which these two configurations have
the same ground state energy depends on the nature of the cutoff. 
For the soft-core interaction potential, 
the crossover between
two ground states takes place when the softening
radius $r_0$ satisfies $R/r_0 \approx 10^{7}$. 
For other forms of regularization, the role of $r_0$ is played by the 
radius of the body of the ring, or the radii of multiple ``cores'' that
constitute the ring.
The maximum angular momentum of a localized self-gravitating system 
scales as 
$$
L_{max} \mathop
{\longrightarrow}_{r_0/R \to 0} \sqrt{\frac{M^3R\ln (R/r_0)}{2\pi}}.
$$ 
In this limit the central core vanishes and the system consist
only  of a ring. 

It would be desirable to check these conclusions by either particle 
simulations or the Mean Field analysis. Unfortunately, at this stage, none of
these seems feasible. As shown in \cite{us} and discussed in
the Introduction, any 
finite-temperature particle simulations will lead to the physically irrelevant
configuration with
the single core sliding along the container wall.
Another obstacle lies in the size of such a computation: To be able see the
crossover between the binary and core-ring system one needs a ring consisting
of $\approx 10^7$ cores, each consisting of at least one particle.
A similar requirement on spatial resolution makes the Mean Field calculations
very challenging too. Indeed, even the existing
calculations with fairly large cutoff \cite{vot,vot1,vot2,vot3}, 
may not have enough mesh
points to resolve the core structure \cite{voterr}.
However, the predicted crossover ratio $R/r_0\approx 10^7$
is not at all astrophysically irrelevant and may
be encountered in the planetary disks and even Saturn rings.
Given that the typical radii of rings around the big planets of the Solar 
system
is of the order of $10^5$ km \cite{sat}, 
the dominant presence of ring particles of the size of 1 cm
and less will make the core-ring configuration thermodynamically  more stable
than the binary one.

The present consideration of the core-ring structure
is rather schematic and at the current level does not allow us to explain the
fine 
structure of rings as gaps and spikes. Neither the Roche limit, which may
set another bound on the existence
of the low-orbit self-gravitating binaries due to the tidal interaction, is
considered. Yet even the present level of modeling of core-ring structures
permits to make important thermodynamical predictions, and 
a distinction between
the existing core-ring models. For example, 
the central core-ring ground state 
configuration considered here
may look similar to the core-ring configuration found
for the system of fermions with fixed angular velocity in \cite{ch2}. 
However, this similarity is only superficial:
The ring observed in \cite{ch2} is formed by particles which could not be
supported gravitationally at the equator of the central core and were shedded
off to the container wall, while in our case the ring is supported only
gravitationally.  
  
Finally, it is interesting to speculate on up to what energy or temperature,
the 
correspondence between the equilibrium and lowest ground-state energy
configurations holds? Indeed, as the energy increases and the mass of the halo
becomes 
comparable or exceeds the combined mass of condensed objects 
(cores and rings), most of the angular momentum is
carried by the halo. This may lead to a reduction and complete
disappearance of a ring.
This scenario looks especially plausible if 
the angular momentum is noticeably smaller than $L_{max}$  so 
a gravitationally-supported  structure with smaller moment of inertia 
than that of an equatorial ring can carry it. In addition to numerical
methods, this
scenario can be analyzed analytically by approximating a halo as a uniform gas,
as done by Chavanis \cite{ch3} in the case of a non-rotating state.
We leave this analysis for future research.

\section{Acknowledgment}
The author is thankful to   R.~J.~Hill, P.-H.~Chavanis, E.~V.~Votyakov, 
M.~Karttunen, and Y.~Kvasnikova for useful comments and discussions 
and gratefully acknowledges the support
of Chilean FONDECYT under the grant 1020052.


\begin{references}
\bibitem{voterr}    The concentration  $\theta =0.02$ considered in
                    \cite{vot} apparently
                    gives for radius of each of two symmetric 
                    cores containing
                    half of all the system particles
                    $R_c=R(3\theta/8\pi)^{1/3} \approx 0.13R$.  
                    Given that only $l_{max}=16$ 
                    spherical harmonics were used, the maximum linear 
                    resolution near the container wall, $2\pi R/l_{max}\approx
                    0.4R$, is clearly
                    insufficient to resolve the cores. For 
                    system without hard-core exclusion considered in 
                    \cite{vot2}, the mere existence of 
                    nonsingular (finite density) cores is a consequence of
                    the softening of
                    interparticle potential due to truncation of
                    harmonic expansion.
\end{references}
\end{document}